\documentclass[preprint,aps,floats,amsmath,amssymb,showpacs]{revtex4}
\usepackage{graphicx}

\begin{document}

\title{Generalised Kramers model}
\author{Vlad Bezuglyy\footnote{Current address : LEMTA, CNRS, 2 avenue de la for\^{e}t de Haye, B.P. 160, 54504
Vandoeuvre Cedex, France\\ E-mail : vladbezuglyy@gmail.com}} \affiliation{Department of Mathematics and Statistics, The Open
University, Walton Hall, Milton Keynes, MK7 6AA, England\\
}

\begin{abstract}
We study a particular generalisation of the classical Kramers model describing Brownian particles in the external potential. The generalised
model includes the stochastic force which is modelled as an additive random noise that depends upon the position of the particle, as well as
time. The stationary solution of the Fokker-Planck equation is analysed in two limits: weak external forcing, where the solution is equivalent
to the increase of the potential compared to the classical model, and strong external forcing, where the solution yields a non-zero probability
flux for the motion in a periodic potential with a broken reflection symmetry.
\end{abstract}


\maketitle

\section{Introduction} This paper addresses the problem of the overdamped motion of independent particles in the external
potential subjected to a random forcing in one spatial dimension.
The model is an extension of the so-called Kramers model
\cite{Kra40} in which a particle at the position $x(t)$ executes
creeping motion according to the following equation of motion:
\begin{equation}
\eta \dot{x}=-\frac{dU}{dx}+f(t).
\end{equation}
Here, $\eta$ is the friction coefficient, $U(x)$ is the potential,
and $f(t)$ is the stochastic force which is usually modelled as a
rapidly fluctuating time-dependent random noise.
%(this can be either a delta-correlated noise or, in more general
%case, a short-correlated random function),
We generalise this model by considering the random force $f(x,t)$
which depends not only upon time, but also upon the position of
the particle. This generalisation is proposed in the same way as
the one discussed for a closely related model of inertial
particles (the Ornstein-Uhlenbeck process \cite{Orn30}) studied
earlier by the author of this paper in collaboration (see
\cite{Arv05,Bez06}). Such a generalisation of the
Ornstein-Uhlenbeck process leads to a number of non-trivial
results: non-Maxwellian stationary distribution of the velocity,
anomalous diffusion of the velocity and position, and \lq
staggered ladder' spectra of the corresponding Fokker-Planck
operator.

The model of Brownian particles in the external potential has a
large number of important applications in physics and chemistry
and below we briefly discuss two of them. First example is a model
of chemical reaction processes, where the position of the particle
represents the reaction coordinate which undergoes a
noise-activated escape process driven by thermal fluctuations
\cite{Han90}. The reaction coordinate is a rather abstract notion
in chemistry characterising the state of a chemical reaction.
Typically, the coordinate wiggles around one of the minima of the
potential energy profile, until a sequence of random `kicks'
induced by thermal fluctuations transports it over the potential
barrier, so that its dynamics can be accurately described by the
motion of the Brownian particle in the external potential.
%An event when the particle overcomes the barrier and
%escapes the minimum corresponds to an advance of the chemical
%reaction.
%It is usually the primary task to calculate the rate at
%the reaction proceeds.

The other interesting application of the Kramers model concerns a concept of the Brownian ratchet, which was originally introduced by Feynman
\cite{Fey66} to illustrate laws of thermodynamics. In its simplest form, the device consists of a ratchet, which resembles a circular saw with
asymmetric teeth, rotating freely in one particular (forward) direction. A pawl is attached to the ratchet, thus preventing it to rotate in the
other (backward) direction. The ratchet is connected to a paddle wheel by a massless frictionless rod and the whole mechanism is immersed in a
thermal bath at a given temperature. It is assumed that the mechanism is so small that the paddle wheel can rotate in response to collisions
with the molecules of the thermal bath, thus rotating back and forth. Because the pawl restricts the backward rotation, the ratchet slowly spins
forward as the molecules hit the paddle-wheel. If a weight is attached to the rod connecting the ratchet and the paddle wheel, it would be
lifted by this forward rotation making the device \lq perpetuum mobile' of the second kind. The contradiction is resolved by noting that the
device must be very small in order to react to individual collisions with the molecules. This means that the pawl itself must be influenced by
the collisions, so that every now and then it would be lifted and fail to prevent the backward rotation. Since both the paddle wheel and the
ratchet are immersed in the same thermal bath, the probability for the pawl to fail is the same as the probability for the ratchet to rotate
forward, so that no net work can be extracted. The analogy with the model of the Brownian particle in the potential is evident. If the position
of the particle represents the angle of rotation of the rod, then the dynamics is periodic and can be split up into two parts: random
fluctuations induced by collisions of the paddle wheel with the molecules and motion in the potential representing the interaction between the
pawl and teeth of the ratchet. The potential in this case is periodic and asymmetric (the so-called \lq sawtooth' potential). The analysis of
the classical model shows that there is no net transport (probability flux) of the Brownian particles moving in a periodic and asymmetric
potential.

In many problems it suffices to know the probability density
function (PDF) of the position of the particle in the steady state
in order to understand all important properties of the Kramers
model. The principal result of this paper is the PDF of the
position of the particle in the generalised model in the limit of
short correlation time of the random force.

We proceed as follows. We start by describing the generalised
model and introducing properties of the stochastic force. In the
limit of short correlation time of the stochastic force the PDF
satisfies the Fokker-Planck equation, which we derive for the
general case. The stationary solution of the Fokker-Planck
equation can be simplified in two asymptotic limits, corresponding
to very large and very small values of the external potential
force. The generalised model in the weak external force limit was
first considered in \cite{Mon08}, where the PDF was found to be
equivalent to a reduction of the potential compared with the
classical Kramers model. Here, a more transparent analysis is used
giving rise to many additional results. We find that in the weak
forcing limit the generalisation leads to an effective increase of
the potential, rather than a decrease derived in \cite{Mon08}. In
the strong forcing limit we find the solution that corresponds to
a non-zero probability flux in the case of the motion in a
periodic potential with a broken reflection symmetry.

\section{Stochastic model} Let us consider a very small particle moving in the potential $U(x)$ and subject to the
stochastic force $f(x,t)$ in one spatial dimension. For a particle
with a negligible mass the velocity is determined by the balance
of the forces acting upon it, so that the equation of motion reads
\begin{equation}
\label{eq:eqm} \eta \dot{x} = -U'(x)+f(x,t),
\end{equation}
where $U'(x)\equiv dU(x)/dx$ is the external potential force. The
random force $f(x,t)$ in (\ref{eq:eqm}) is assumed to be a
stationary and translationally invariant Gaussian process with
zero mean and correlation function
\begin{equation}
\langle f(x,t)f(x',t')\rangle=C(x-x',t-t'),
\end{equation}
where angular brackets denote average over noise realisations
throughout. The noise has a typical magnitude $\sigma$,
correlation length $\xi$, and correlation time $\tau$. We assume
that the correlation function is smooth and sufficiently
differentiable and decays rapidly for $|x|>\xi$ and $|t|>\tau$. In
the absence of the external potential the particle is not bounded
and diffuses, so that the mean square displacement is given by
\mbox{$\langle [x(t)-x(0)]^2\rangle \sim 2D_xt$} with a diffusion
constant \mbox{$D_x\sim \sigma^2 \tau/\eta^2$} for $t\gg \tau$.
Relaxation towards a statistically stationary state is associated
with the action of the potential. The corresponding relaxation
time $T$ depends upon particular properties of the potential, as
well as properties of the random force, but in the general case it
cannot be determined explicitly.

\section{Fokker-Planck equation}
If the correlation time of the random force is sufficiently short
(that is $\tau\ll T$), it is possible to define a time scale
$\delta t$ at which the stochastic force fluctuates appreciably,
while the change of the dynamical variable $x(t)$ is negligible on
the length scale of the potential, $L$. Integrating the equation
of motion \eqref{eq:eqm} over the time period $\delta t$ we obtain
\begin{equation}
\label{eq:lang} \delta x \equiv x(t_0 + \delta t)-x(t_0) =
-\frac{U'(x)}{\eta}\delta t + \frac{1}{\eta}\int
_{t_0}^{t_0+\delta t} dt \ f(x(t),t).
\end{equation}
%
%
%It is also assumed that the potential varies slowly on the length
%scale of the distance travelled by the particle in $\delta t$ due
%to the random noise. If we denote the dimensionless measure of the
%strength of the noise by $\chi = \sigma \tau/(\eta \xi)$, then the
%Langevin approach is valid for $\chi \lambda \ll 1$.
Following the standard procedure (see, e.g. \cite{Kam81}), we
write the Fokker-Planck equation for the probability density
function $P(x,t)$ for the stochastic model given by
Eq.~\eqref{eq:eqm} in the limit of short correlation time of the
random force:
\begin{equation}
\label{eq:fp} \frac{\partial P(x,t)}{\partial
t}=-\frac{\partial}{\partial x}
[v(x)P(x,t)]+\frac{\partial^2}{\partial x^2}[D(x)P(x,t)].
\end{equation}
Here, $v(x)$ is the drift velocity and $D(x)$ is the diffusion
coefficient defined via the increment $\delta x$ as follows:
\begin{eqnarray}
v(x)&=&\frac{\langle \delta x \rangle}{\delta t},\nonumber \\
D(x)&=&\frac{\langle \delta x^2 \rangle}{2\delta t}.
\end{eqnarray}
In the following sections we use stationary and translationally
invariant properties of the noise and set $t_0=0$ and $x(t_0)=0$
in Eq.~\eqref{eq:lang} for calculating statistical properties of
$\delta x$. Using Eq.~\eqref{eq:lang} we obtain
\begin{eqnarray}
v(x)&=&-\frac{U'(x)}{\eta}+\frac{1}{\delta t}\frac{1}{\eta}\int
_{0}^{\delta t} dt \ \langle f(x(t),t)\rangle ,\nonumber \\
D(x)&=&\frac{1}{2\delta t}\frac{1}{\eta^2}\left\langle \left[\int
_{0}^{\delta t} dt \ f(x(t),t)\right]^2\right \rangle.
\end{eqnarray}

We are interested in the stationary solution of Eq.~\eqref{eq:fp}
satisfying $\partial _t P(x,t)=0$. It is found by solving the
differential equation
\begin{equation}
\label{eq:current} -v(x)P_0(x)+\frac{\partial}{\partial
x}[D(x)P_0(x)]=-J_0,
\end{equation}
where the stationary probability flux $J_0$ is determined from the
boundary conditions. The solution of Eq.~\eqref{eq:current} can be
readily written as
\begin{equation}
\label{eq:gen} P_0(x)=Z(x)\left[N-J_0\int _0 ^x dy \ D^{-1}(y)
Z^{-1}(y)\right],
\end{equation}
where
\begin{equation}
Z(x)={\rm exp}\left[\int _0^x  \ dy \frac{v(y)-D'(y)}{D(y)}\right]
\end{equation}
and $N$ is the normalisation constant. We remark that in the case
of a periodic potential $P_0(x)$ is normalised in the periodicity
interval. The rest of the paper is concerned with simplifying the
solution \eqref{eq:gen} in two asymptotic limits corresponding to
very large and very small values of the external force $U'(x)$.

It is not typical to have a non-zero flux $J_0$ in systems that
are in thermal equilibrium. The cases where the transport can be
introduced by different mechanisms are of great interest. Feynman
considered the case where the ratchet and the paddle-wheel are
immersed in separate thermal baths at different temperatures. In
this case, the transport is induced by the gradient of the
temperature. The transport in the Kramers model may also be
induced by an addition of another driving force that can be
constant \cite{Rei02} or a function of time \cite{Mag93}. We also
remark that the Fokker-Planck equation with the state-dependent
diffusion coefficient was studied before in \cite{But1987, Kam88},
where the transport in a symmetric periodic potential is a
consequence of the non-uniform intensity of the stochastic force
modelled as a multiplicative noise, i.e. $f(x,t)=g(x)h(t)$, where
$g(x)$ is periodic and $h(t)$ is a rapidly fluctuating random
noise. In this paper we show that it is possible to obtain a
non-zero flux even for a model where the noise is additive and has
translationally invariant statistics.

We conclude this section by discussing conditions and limits of
validity of the Fokker-Planck equation for our model. The question
of the validity of the Fokker-Planck approach is rather hard to
discuss in precise terms for the problems which involve spatial
dependence of the additive noise. The important quantity in this
case is the effective correlation time of the stochastic force,
i.e. how rapidly the force experienced by the moving particle
de-correlates. It is evident that the additional correlation in
space may only decrease this effective correlation time. The
Fokker-Planck approach relies mainly on two conditions: short
correlation time of the stochastic force and small change of the
dynamical variable in $\delta t$. The first condition has already
been mentioned earlier and reads $\tau/T\ll 1$, where $T$ is the
relaxation time. %This conditions appears to be sufficient for the
%theory to work in the view of the discussion above.
As for the small increment, the obvious condition would be $\sigma
\tau \ll L$. Again, this condition is only approximate, since the
effective correlation time is not known explicitly.

\section{Weak external force limit} 
\begin{figure}[t]
\centerline{\includegraphics[width=14cm]{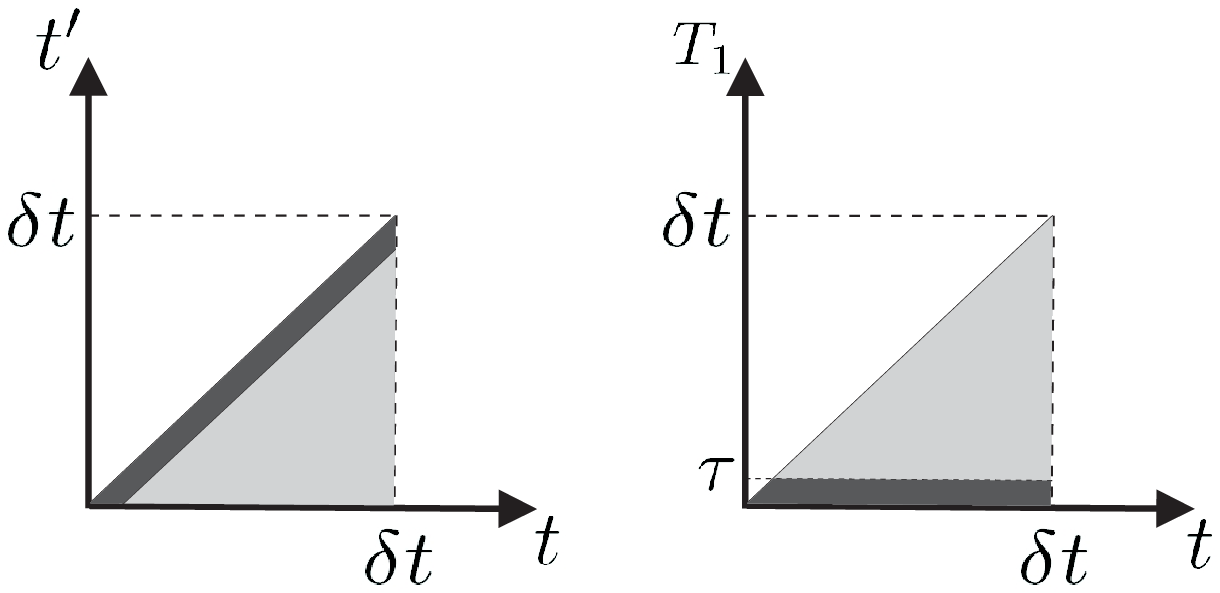}}
\caption{\label{fig:int} Shows the transformation of variables
from Eqs.~\eqref{eq:i1} (left) to \eqref{eq:i2} (right). Light
grey areas indicate the domain of integration and dark grey areas
indicate regions where the integrand is significant.}
\end{figure}

Let us consider the increment
$\delta x$ in the limit when the motion of the particle is
dominated by the stochastic force. %In what follows we shall
%neglect terms higher than $O[U'(x)]$.
First, we introduce some additional notation:
\begin{eqnarray}
\label{eq:aux}
s(x,t)&=&-\frac{1}{\eta}U'(x)t,\nonumber \\
x^{(0)}(t)&=&\frac{1}{\eta}\int _0^t dt' \ f(x(t'),t').
\end{eqnarray}
Using this we can write the increment from Eq.~\eqref{eq:lang} as
follows:
\begin{equation}
\delta x = s(x,\delta t)+x^{(0)}(\delta t).
\end{equation}
Expanding in the series the stochastic force about $x=x^{(0)}(t)$
we obtain
\begin{equation}
\label{eq:dxs} \delta x = s(x,\delta t)+\frac{1}{\eta}\int _0
^{\delta t} dt \ f(x^{(0)}(t),t)-\frac{U'(x)}{\eta^2}\int _0
^{\delta t} dt \ t \ \frac{\partial f(x^{(0)}(t),t)}{\partial
x}+O[U'(x)]^2.
\end{equation}
Averaging this expression we obtain
\begin{equation}
\langle \delta x \rangle \approx s(x,\delta t)+\frac{1}{\eta}\int
_0 ^{\delta t} dt \ \langle f(x^{(0)}(t),t)\rangle
-\frac{U'(x)}{\eta^2}\int _0 ^{\delta t} dt \ t \ \left \langle
\frac{\partial f(x^{(0)}(t),t)}{\partial x}\right \rangle.
\end{equation}
We now simplify the problem by considering the case when the
spatial dependence of the random force is weak or, equivalently,
when the correlation length is sufficiently large. Let us
introduce a quantity which measures a distance travelled by the
particle due to the random force in the correlation time relative
to the correlation length:
\begin{equation}
{\rm Ku}=\frac{\sigma \tau}{\xi \eta}.
\end{equation}
We term this parameter the Kubo number. It has been used before in
the similar context of motion of inertial particles (see e.g.
\cite{Wil1}). We remark that the classical Kramers model
corresponds to ${\rm Ku}=0$. When the Kubo number is small, we can
write the firs moment of $\delta x$ by expanding the stochastic
force further:
\begin{eqnarray}
\langle \delta x \rangle &\approx & s(x,\delta
t)+\frac{1}{\eta}\int _0 ^{\delta t} dt \  \left \langle f(0,t) +
\frac{\partial f(0,t)}{\partial x} x^{(0)}(t)\right
\rangle\nonumber\\ &-&\frac{U'(x)}{\eta^2}\int _0 ^{\delta t} dt \
t \ \left \langle\left [ \frac{\partial f(0,t)}{\partial
x}+\frac{\partial^2 f(0,t)}{\partial x^2}x^{(0)}(t) \right]\right
\rangle.
\end{eqnarray}
From the properties of the random force we have $\langle
f(0,t)\rangle =0$ and $\langle \partial _x f(0,t)\rangle =0$, and
using the definition of $x^{(0)}(t)$ we obtain
\begin{eqnarray}
\label{eq:exp1} \langle \delta x \rangle &\approx & s(x,\delta
t)+\frac{1}{\eta^2}\int _0 ^{\delta t} dt \int _0 ^t dt' \
\left\langle \frac{\partial f(0,t)}{\partial x}f(x(t'),t') \right
\rangle \nonumber \\ &-&\frac{U'(x)}{\eta^3}\int _0 ^{\delta t} dt
\int _0 ^t dt' \ t \ \left \langle \frac{\partial^2
f(0,t)}{\partial x^2}f(x(t'),t') \right \rangle.
\end{eqnarray}
Next, we expand $f(x(t'),t')=f(0,t')+\partial _x
f(0,t')[s(x,t')+x^{(0)}(t')]$. We note that $\langle
f(0,t_1)\partial _x f(0,t_2)\rangle=0$ for any $t_1$ and $t_2$,
and after dropping terms of order higher than $U'(x)$ we obtain
\begin{eqnarray}
\langle \delta x \rangle & \approx &   s(x,\delta
t)+\frac{1}{\eta^2}\int _0 ^{\delta t} dt \int _0 ^t dt' \ s(x,t')
\left\langle \frac{\partial f(0,t)}{\partial x} \frac{\partial
f(0,t')}{\partial x} \right \rangle\nonumber \\
&-&\frac{U'(x)}{\eta^3}\int _0 ^{\delta t} dt \int _0 ^t dt' \ t \
\left \langle \frac{\partial^2 f(0,t)}{\partial x^2}f(0,t') \right
\rangle.
\end{eqnarray}
For the two-point correlation function in this expression we use
the following identities which hold for any stationary Gaussian
noise:
\begin{eqnarray}
\left\langle \frac{\partial f(0,t)}{\partial x} \frac{\partial
f(0,t')}{\partial x} \right \rangle &=& -\frac{\partial^2
C(0,t-t')}{\partial x^2},\nonumber\\
\left\langle \frac{\partial^2 f(0,t)}{\partial x^2} f(0,t') \right
\rangle &=& \frac{\partial^2 C(0,t-t')}{\partial x^2}.
\end{eqnarray}
Using this we obtain
\begin{eqnarray}
\langle \delta x \rangle &\approx & s(x,\delta
t)+\frac{U'(x)}{\eta^3}\int _0 ^{\delta t} dt \int _0 ^t dt' \ t'
\ \frac{\partial^2 C(0,t-t')}{\partial x^2}\nonumber \\
&-&\frac{U'(x)}{\eta^3}\int _0 ^{\delta t} dt \int _0 ^t dt' \ t \
\frac{\partial^2 C(0,t-t')}{\partial x^2} \nonumber \\
&=& s(x,\delta t)-\frac{U'(x)}{\eta^3}\int _0 ^{\delta t} dt \int
_0 ^t dt' \ (t-t') \ \frac{\partial^2 C(0,t-t')}{\partial x^2}.
\end{eqnarray}
The integrand in the last term depends only upon $t-t'$ and is
therefore linear in $\delta t$. In the remaining part of the paper
we shall deal with similar double and quadruple integrals, so now
we discuss the last term in more details. Let us consider a double
integral
\begin{equation}
\label{eq:i1} Q= \int _0 ^{\delta t} dt \int _0 ^t dt' \ (t-t') \
\frac{\partial^2 C(0,t-t')}{\partial x^2}.
\end{equation}
We denote $T_1=t-t'$ and obtain
\begin{equation}
\label{eq:i2} Q= \int _0 ^{\delta t} dt \int _0 ^t dT_1 \ T_1 \
\frac{\partial^2 C(0,T_1)}{\partial x^2}.
\end{equation}
In Fig.~\ref{fig:int} we illustrate this transformation of
variables. For $\delta t\gg \tau$ the integrand is significant
around $T_1=0$ and decreases rapidly for $T_1$ increasing. Thus,
if we integrate for $T_1$ from $0$ to $\infty$, we would only make
a small error of order $\tau^2$. Using this we may write
\begin{equation}
Q\approx \int _0 ^{\delta t} dt \int _0 ^{\infty} dT_1 \ T_1 \
\frac{\partial^2 C(0,T_1)}{\partial x^2}=\delta t \int _0
^{\infty} dT_1 \ T_1 \ \frac{\partial^2 C(0,T_1)}{\partial x^2}.
\end{equation}
Assuming that the last integral is convergent we obtain
\begin{equation}
Q\approx c\delta t, \quad c=\int _0 ^{\infty} dT_1 \ T_1 \
\frac{\partial^2 C(0,T_1)}{\partial x^2}.
\end{equation}

We return to the calculation of $\langle \delta x \rangle$ and
obtain
\begin{equation}
\langle \delta x \rangle = s(x,\delta
t)-\frac{U'(x)}{\eta^3}\delta t \int _0 ^{\infty} dt \ t \
\frac{\partial^2 C(0,t)}{\partial x^2}.
\end{equation}
The drift velocity then reads
\begin{equation}
\label{eq:drift1} v(x) = -\frac{U'(x)}{\eta}(1-\alpha),
\end{equation}
where
\begin{equation}
\label{eq:alpha} \alpha=-\frac{1}{\eta^2}\int _0 ^{\infty} dt \ t
\ \frac{\partial^2 C(0,t)}{\partial x^2}.
\end{equation}
The sign of $\alpha$ can be deduced as follows. If we can write the correlation function in the form $C(x,t)=C_x(x)C_t(t)$, where $C_t(t)>0$,
then the sign of $\alpha$ is determined by the sign of $C''_x(0)$. If the random force de-correlates as $x$ increases, then $x=0$ is a local
maximum of $C_x(x)$. Providing that the second derivative exists, it follows that $C''_x(0)<0$ and, consequently, $\alpha>0$. Furthermore, we
have $\partial _{xx} C(0,0) \sim \sigma^2/\xi^2$, and therefore $\alpha \sim {\rm Ku}^2$. We remark that if we keep expanding the stochastic
force further in Eq.~\eqref{eq:exp1}, we would obtain terms of order higher than ${\rm Ku}^2$.

\begin{figure}[t]
\centerline{\includegraphics[width=14cm]{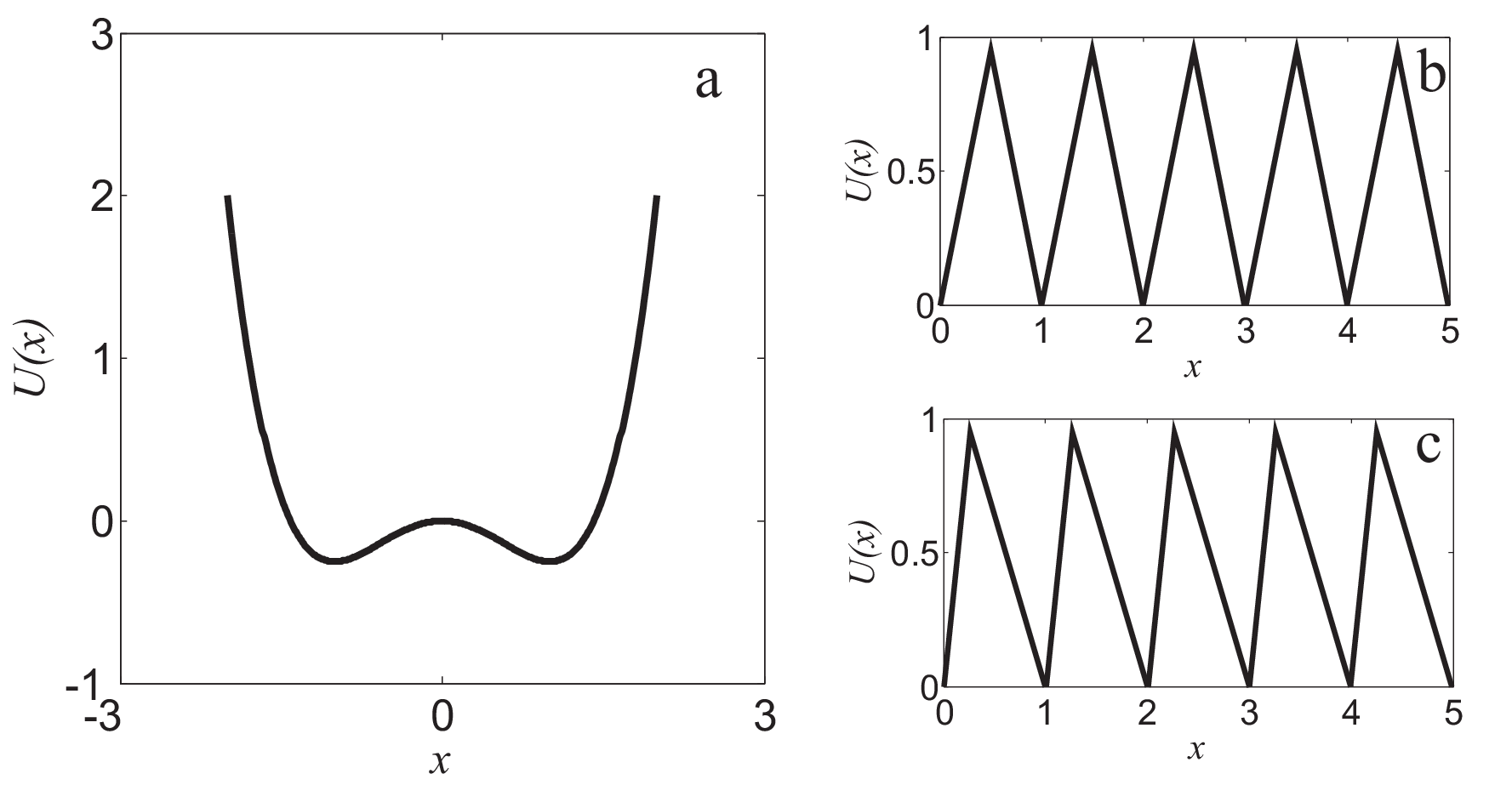}}
\caption{\label{fig:1a} Shows three examples of the external
potential: symmetric double-well potential ({\bf a}), periodic
linear piece-wise potential ({\bf b}), and periodic linear
piece-wise potential with a broken reflection symmetry ({\bf c}).}
\end{figure}

We now calculate $\langle \delta x^2 \rangle$ and the diffusion coefficient. After squaring and averaging Eq.~\eqref{eq:dxs} we obtain
\begin{eqnarray}
\langle \delta x ^2\rangle &=&\frac{1}{\eta^2}\left\langle \left
[\int _0 ^{\delta t} dt \ f(x^{(0)}(t),t) \right]^2\right
\rangle\nonumber \\ &-& \frac{2U'(x)}{\eta^3} \int _0 ^{\delta t}
dt_1 \int _0 ^{\delta t} dt_2 \ t_2 \ \left\langle
f(x^{(0)}(t_1),t_1)\frac{\partial f(x^{(0)}(t_2),t_2)}{\partial
x}\right \rangle+O[U'(x)]^2.
\end{eqnarray}
The second term on the right hand side in this expression is at
least $O(\delta t)^2$. This becomes obvious if we notice that the
correlation function in the integrand depends on $t_1-t_2$, but
due to the factor $t_2$ the whole integrand cannot be expressed as
a function of $t_1-t_2$ only. The diffusion coefficient is
therefore given by
\begin{equation}
D=\frac{1}{2\delta t} \frac{1}{\eta^2}\left\langle \left [\int _0
^{\delta t} dt \ f(x^{(0)}(t),t) \right]^2\right \rangle.
\end{equation}
If we proceed to expand the stochastic force further, we would
obtain terms which are at least $O[U'(x)]^2$. We therefore
conclude that the diffusion coefficient in this case is constant
and is the same as in the model of free diffusion given by the
equation
\begin{equation}
\dot{x}=\frac{f(x,t)}{\eta}.
\end{equation}
Let us now consider a case of small Kubo number similarly to the
calculation of the drift velocity. We shall consider this case as
a separate problem and discuss it in the appendix. We obtain that
in the limit of small ${\rm Ku}$ (or small $\alpha$) the diffusion
constant is given by
\begin{equation}
\label{eq:diff1} D=D_0[1-(2+\gamma)\alpha],
\end{equation}
where $D_0$ is the diffusion constant for the model in the absence
of the spatial correlation corresponding to ${\rm Ku}=0$. It is
given by
\begin{equation}
D_0=\frac{1}{2\eta^2} \int _{-\infty}^{\infty} dt \ C(0,t).
\end{equation}
The factor $\gamma>0$ is given by
\begin{equation}
\gamma =\frac{1}{2\alpha D_0\eta^4} \int _{-\infty}^{\infty} dt_1
\frac{\partial^2 C(0,t_1)}{\partial x^2} \int _{0}^{t_1} dt_2 \int
_{t_2}^{\infty} dt_3 \ C(0,t_3).
\end{equation}
Thus, the diffusion constant is reduced by the factor
$1-\alpha(2+\gamma)$ compared to the case of ${\rm Ku}=0$. Using
Eqs.~\eqref{eq:drift1} and \eqref{eq:diff1} we obtain the solution
of the Fokker-Planck equation in the weak forcing limit
corresponding to small ${\rm Ku}$:
\begin{equation}
\label{eq:sol1} P_0(x)=Y(x)\left[N-\frac{J_0}{D_0}\int _0 ^x dy \
Y^{-1}(y)\right],
\end{equation}
where
\begin{equation}
Y(x)={\rm exp}\left[-\frac{U(x)[1+\alpha(1+\gamma)]}{\eta
D_0}\right ].
\end{equation}
We now concentrate on the form of the solution \eqref{eq:sol1} for
particular choices of the potential illustrated in
Fig.~\ref{fig:1a}. First example is a symmetric double-well
potential used in modelling two-way chemical reactions, and the
other is a periodic potential with period $L$. For the double-well
potential illustrated in Fig.~\ref{fig:1a}a the natural boundary
conditions are applied \cite{Ris99}:
\begin{equation}
P(\infty) = P(-\infty) = 0.
\end{equation}
Such a potential does not allow the particles to escape to
infinity, so that we expect that the probability flux vanishes. We
note that $Y(x)$ goes to zero for very large $x$ and the second
term in the brackets multiplied by $Y(x)$ approaches a non-zero
constant. Thus, the boundary conditions are satisfied only when
$J_0=0$.

For the periodic potential, if we require that $P_0(x)$ is bounded
for the increasing $x$, it follows that $P_0(x)$ is periodic
\cite{Ris99}. We use $U(x+L)=U(x)$ to obtain $Y(x+L)=Y(x)$ and
therefore the condition of periodicity reads
\begin{equation}
P_0(x+L)=P_0(x)-\frac{J_0}{D_0}\int _x ^{x+L} dy \ Y(x)Y^{-1}(y).
\end{equation}
The integral in the last term is non-zero, therefore we again put
$J_0=0$ to satisfy the boundary conditions. The important
consequence of this result is that the flux vanishes regardless of
the shape of the periodic potential. In the studies of Brownian
ratchets it is often assumed that the periodic potential has an
asymmetric form (such as the \lq sawtooth' potential illustrated
in Fig.~\ref{fig:1a}c), so that the particles are expected to
favour the slope with a smaller inclination to escape the
potential minimum. The result shows, however, that the probability
flux vanishes, which agrees with the discussion of the Brownian
ratchet in the introduction.

We conclude that in both examples the solution in the weak
external force limit is given by
\begin{equation}
\label{eq:p01} P_0(x)= N{\rm exp}\left[
-\frac{U(x)[1+\alpha(1+\gamma)]}{\eta D_0} \right].
\end{equation}
This is the Maxwellian density with the potential increased by the
factor $1+\alpha(1+\gamma)$ compared with the classical Kramers
model, which corresponds to $\alpha=0$. The solution is consistent
with the idea that in the presence of spatial correlations the
noise experienced by the moving particle de-correlates more
rapidly than for the case of an infinite correlation length in the
classical Kramers model. This means that the particle experiences
more uncorrelated kicks along its trajectory decreasing the
probability to travel far against the systematic force $-U'(x)$.
Therefore, we expect to see the density function becoming sharper
around the minima of the potential as the correlation length
decreases. Our result differs from the one obtained in
\cite{Mon08}, where the effective decrease of the potential in the
solution is attributed to the reduction of the drift velocity
given by Eq.~\eqref{eq:drift1}, but the corresponding reduction of
the diffusion coefficient is not considered.

We remark that in the general case, when the potential force is
weak, the drift reduction remains linear in $U'(x)$ and the
diffusion coefficient remains constant, even when ${\rm Ku}$ is
not small. The actual values of $\alpha$ and $\gamma$ in the case
of arbitrary Kubo number are not known, but the density still
remains Maxwellian around stagnation points of the potential,
provided that the Fokker-Planck approach remains valid.

\section{Strong external force limit} \label{sec}
In this section we analyse the limit when the motion of the
particle is dominated by the external potential force. In this
case we can expand the stochastic force in the series about
$x=s(x,t)$. The increment $\delta x$ in this case reads
\begin{equation}
\delta x \approx s(x,\delta t)+\frac{1}{\eta}\int _0 ^{\delta t}
dt \ f(s(x,t),t)+\frac{1}{\eta}\int _0^{\delta t} dt \
\frac{\partial f(s(x,t),t)}{\partial x}x^{(0)}(t).
\end{equation}
We first calculate $\langle \delta x^2\rangle$. The term
$[s(x,\delta t)]^2$ is obviously of order $\delta t^2$ and so is
the mixed product of $s(x,\delta t)$ and the integral terms in the
expression above. The rest of the terms require some careful
considerations. We have
\begin{eqnarray}
\label{eq:dx1} \langle \delta x^2 \rangle &=&
\frac{1}{\eta^2}\left \langle \left [ \int _0 ^{\delta t} dt \
f(s(x,t),t)\right]^2\right \rangle +\frac{1}{\eta^2}\left \langle
\left [ \int _0^{\delta t} dt \ \frac{\partial
f(s(x,t),t)}{\partial x}x^{(0)}(t)\right]^2\right
\rangle\nonumber \\
&+&\frac{2}{\eta^2} \int _0^{\delta t} dt_1 \int _0^{\delta t}
dt_2 \ \left \langle f(s(x,t_1),t_1) \frac{\partial
f(s(x,t_2),t_2)}{\partial x}x^{(0)}(t_2)\right \rangle.
\end{eqnarray}
For the first term we obtain
\begin{eqnarray}
\frac{1}{\eta^2}\left \langle \left [ \int _0 ^{\delta t} dt \
f(s(x,t),t)\right]^2\right \rangle&=&\frac{1}{\eta^2} \int
_0^{\delta t} dt_1 \int _0^{\delta t} dt_2 \ \langle
f(s(x,t_1),t_1)
f(s(x,t_2),t_2)\rangle \nonumber \\
&=& \frac{1}{\eta^2} \int _0^{\delta t} dt_1 \int _0^{\delta t}
dt_2 \ C(s(x,t_1-t_2),t_1-t_2).
\end{eqnarray}
The integrand in the last expression depends only on $t_1-t_2$ and
is therefore of order $\delta t$ when $\delta t\gg \tau$.
Similarly to the cases considered in the previous section (see
Eq.~\eqref{eq:i1}) we obtain
\begin{equation}
\frac{1}{\eta^2} \int _0^{\delta t} dt_1 \int _0^{\delta t} dt_2 \
\langle f(s(x,t_1),t_1) f(s(x,t_2),t_2)\rangle = \frac{\delta
t}{\eta^2} \int _{-\infty}^{\infty} dt \ C(s(x,t),t).
\end{equation}
We now proceed a step further and calculate this term expanding
for large $U'(x)$. Using the definition of $s(x,t)$ in
Eq.~\eqref{eq:aux} we can write this by changing the variable from
$t$ to $z\equiv s(x,t)$
\begin{eqnarray}
\label{eq:t1} \frac{\delta t}{\eta^2} \int _{-\infty}^{\infty} dt
\ C(s(x,t),t)&=&\frac{\delta t}{|U'(x)|\eta} \int
_{-\infty}^{\infty} dz \ C(z,-z\eta/U'(x))\nonumber \\
&=&\frac{\delta t}{|U'(x)|\eta } \int _{-\infty}^{\infty} dz \
C(z,0)+O[U'(x)]^{-2}.
\end{eqnarray}
The modulus sign is used to ensure that the expression remains
positive. Thus, in the limit of strong external forcing the first
term in $\langle \delta x^2 \rangle $ is inversely proportional to
$|U'(x)|$. Now we return to the starting point
(Eq.~\eqref{eq:dx1}) and consider for instance the term
\begin{eqnarray}
&&\frac{1}{\eta^2}\left \langle \left [ \int _0^{\delta t} dt \
\frac{\partial f(s(x,t),t)}{\partial x}x^{(0)}(t)\right]^2\right
\rangle =\frac{1}{\eta^2} \int _0^{\delta t} dt_1 \int _0^{\delta
t} dt_2 \int _0^{t_1} dt' \int _0^{t_2} dt'' \nonumber \\
& \times &\left \langle \frac{\partial f(s(x,t_1),t_1)}{\partial
x} \frac{\partial f(s(x,t_2),t_2)}{\partial x} f(s(x,t'),t')
f(s(x,t''),t'')\right \rangle.
\end{eqnarray}
The four-point correlation function for a Gaussian random process
can be expressed as the sum of all possible non-repeating
combinations of products of two-point correlation functions. A
typical combination in this case may look as follows:
\begin{eqnarray}
&&\left \langle \frac{\partial f(s(x,t_1),t_1)}{\partial x}
\frac{\partial f(s(x,t_2),t_2)}{\partial x} \right \rangle \langle
f(s(x,t'),t') f(s(x,t''),t'')\rangle \nonumber
\\ &=&-\frac{\partial^2 C(s(x,t_1-t_2),t_1-t_2)}{\partial
x^2}C(s(x,t'-t''),t'-t'').
\end{eqnarray}
If we proceed in the same way as for the previous term expanding
for large $U'(x)$, each of the factors would contribute at least
$[U'(x)]^{-1}$, so that the overall contribution would be of order
$[U'(x)]^{-2}$, and therefore may be neglected. Similarly, the
remaining term in Eq.~\eqref{eq:dx1} may also be neglected. We
conclude that the diffusion coefficient is determined by
Eq.~\eqref{eq:t1} and reads
\begin{equation}
D(x)=\frac{1}{2|U'(x)|\eta } \int _{-\infty}^{\infty} dz \ C(z,0).
\end{equation}
We rewrite this as follows:
\begin{equation}
\label{eq:d1} D(x)=\frac{D_{\infty}}{|U'(x)|\eta}, \quad
D_{\infty}=\frac{1}{2} \int _{-\infty}^{\infty} dz \ C(z,0).
\end{equation}

We now consider $\langle \delta x \rangle$:
\begin{equation}
\langle \delta x \rangle\approx s(x,\delta t)+\frac{1}{\eta}\int
_0^{\delta t} dt \ \langle f(s(x,t),t)\rangle +\frac{1}{\eta}\int
_0^{\delta t} dt \ \left \langle \frac{\partial
f(s(x,t),t)}{\partial x}x^{(0)}(t)\right\rangle.
\end{equation}
Here, the second term vanishes because effectively the average is
taken over a deterministic trajectory, since the potential is
assumed to be varying slowly. For the second term we obtain
\begin{equation}
\frac{1}{\eta}\int _0^{\delta t} dt \ \left\langle \frac{\partial
f(s(x,t),t)}{\partial x}x^{(0)}(t)\right \rangle\approx
\frac{1}{\eta^2}\int _0^{\delta t} dt \int _0^t dt_1 \ \left
\langle \frac{\partial f(s(x,t),t)}{\partial
x}f(s(x,t_1),t_1)\right\rangle
\end{equation}
neglecting terms of higher orders in $x^{(0)}(t)$. We then obtain
\begin{equation}
\frac{1}{\eta^2}\int _0^{\delta t} dt \int _0^t dt_1 \ \left
\langle \frac{\partial f(s(x,t),t)}{\partial
x}f(s(x,t_1),t_1)\right\rangle\approx \frac{\delta t}{\eta^2}\int
_0^{\infty} dt \ \frac{\partial C(s(x,t),t)}{\partial x}.
\end{equation}
If we proceed further and expand this expression for strong
external force, we obtain the term which is inverse proportional
to $U'(x)$, similarly to the calculation of $\langle \delta x^2
\rangle$. Since $s(x,\delta t)\sim U'(x)$, we therefore conclude
that the first moment of $\delta x$ in the limit of strong
external force reads
\begin{equation}
\langle \delta x\rangle=s(x,\delta t)+O[U'(x)]^{-1}.
\end{equation}
The drift velocity is therefore given by
\begin{equation}
\label{eq: v2} v(x)=-\frac{U'(x)}{\eta}.
\end{equation}
Substituting \eqref{eq:d1} and \eqref{eq: v2} into \eqref{eq:gen}
we obtain the solution in the strong external force limit:
\begin{equation}
P_0(x)=|U'(x)|{\rm
e}^{-I(x)}\left[N-\frac{J_0\eta}{D_{\infty}}\int _0^x dy \ {\rm
e}^{I(y)}\right],
\end{equation}
where
\begin{equation}
\label{eq:ix} I(x)=\frac{1}{D_{\infty}}\int _0^x dy \
|U'(y)|U'(y).
\end{equation}
For the non-periodic potential, if $P_0(\pm \infty)=0$ we can
again show that $J_0=0$. We note that $I(x)$ diverges for large
$x$, whereas the integral term in the brackets multiplied by ${\rm
exp}[-I(x)]$ converges to a constant for large $x$. The solution
corresponding to $J_0=0$ is given by
\begin{equation}
\label{eq:p03} P_0(x)=N|U'(x)|{\rm e}^{-I(x)}.
\end{equation}
For the case of a periodic potential with the period $L$ we find
$J_0$ by writing $P(L)=P(0)$ as
\begin{equation}
|U'(L)|{\rm e}^{-I(L)}\left[N-\frac{J_0\eta}{D_{\infty}}\int _0^L
dy \ {\rm e}^{I(y)}\right]=N|U'(0)|.
\end{equation}
Using $U'(0)=U'(L)$ we obtain
\begin{equation}
\label{eq:j} J_0=\frac{N D_{\infty}[1-{\rm e}^{I(L)}]}{\eta \int
_0^L dy \ {\rm e}^{I(y)}}.
\end{equation}
We note that $I(x-L)=I(x)-I(L)$ for the periodic potential and
thus the solution in the strong external force limit can be
written in the following compact form:
\begin{equation}
\label{eq:p02} P_0(x)=N|U'(x)|{\rm e}^{-I(x)}\int _x^{x+L} dy \
{\rm e}^{I(y)}.
\end{equation}

For the periodic potential we have obtained a peculiar result: if
$I(L)\neq 0$ the solution of the Fokker-Planck equation
corresponds to a non-zero probability flux in the stationary
state. For any periodic potential integrating $U'(x)$ in the
periodicity interval gives 0. Thus, if the expression for $I(x)$
contained only $U'(x)$, the flux would vanish. Because the
integrand in Eq.~\eqref{eq:ix} is quadratic in $U'(x)$, the sign
of $I(L)$ is determined by the sign of the steepest of two slopes
of the potential, if we consider a case of the potential with a
single minimum in the periodicity interval. Thus, if the periodic
potential is symmetric (such as the one in Fig.~\ref{fig:1a}b),
then $I(L)=0$ and $J_0$ vanishes. Conversely, for a \lq sawtooth'
potential with a broken reflection symmetry (Fig.~\ref{fig:1a}c),
$I(L)\neq 0$ and the solution of the Fokker-Planck equation
corresponds to a non-zero probability flux.

\section{Numerical simulations and discussion}
We perform a number of numerical experiments in order to
illustrate our analytical results. Numerical simulations are done
by integrating the original equation of motion \eqref{eq:eqm}
using a small time step (typically about $\tau/50$). In the
simulations we use the following correlation function of the
random force:
\begin{equation}
\label{eq:corr1} C(x,t)=\sigma^2{\rm
exp}\left(-\frac{x^2}{2\xi^2}-\frac{t^2}{2\tau^2}\right).
\end{equation}
We use two different types of the potential corresponding to the
examples given in sections IV and V: an asymmetric periodic
potential $U(x)=(V_0 L/2\pi) [ \sin (2\pi x/L)+ k\sin (4\pi x/L)]$
and a non-periodic double-well potential $U(x)=x^4/4-x^2/2$. The
relaxation time $T$ is of order unity in all simulations (as
judged from the plot $\langle x^2(t)\rangle$ versus $t$), so that
the Fokker-Planck approach is valid for the values of $\tau$
typically smaller than $10^{-1}$.

\subsection{Free diffusion}
\begin{figure}[t]
\centerline{\includegraphics[width=8cm]{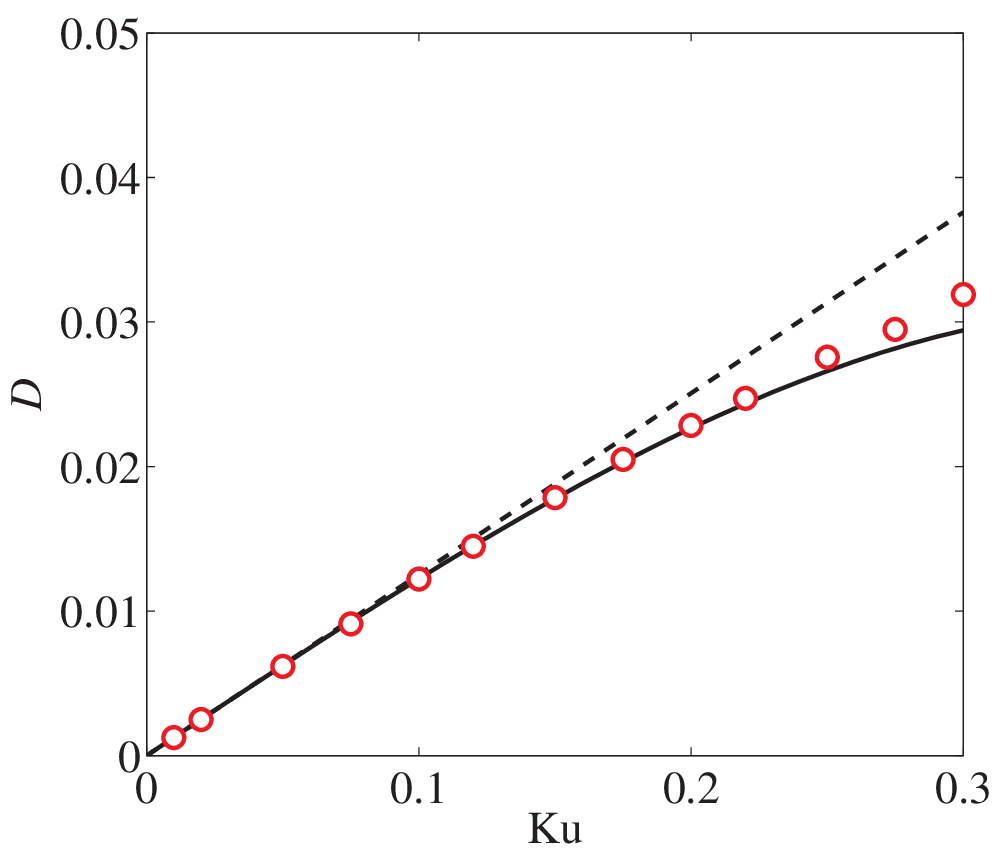}}
\caption{\label{fig:diff} Shows the diffusion constant in the
model of free diffusion. The data from the numerical simulations
(circles) are compared with Eq.~\eqref{eq:diff1} (solid line).
Also shown is the diffusion constant $D_0$ corresponding to
$\xi\rightarrow\infty$ (dashed line). In the simulations we fix
$\sigma=1.0$, $\xi=0.1$, and $\eta=1.0$ and vary $\tau$.}
\end{figure}

We start by illustrating the reduction of the diffusion constant
in the model of free diffusion ($U(x)=0$) in the limit of small
Kubo number. For the correlation function given by
Eq.~\eqref{eq:corr1} we obtain
\begin{eqnarray}
D_0&=&\sqrt{\frac{\pi}{2}}\frac{\sigma^2}{\eta^2}\tau,\nonumber\\
\alpha&=&\frac{\sigma^2\tau^2}{\eta^2\xi^2}={\rm Ku^2},\nonumber\\
\gamma &=& \sqrt{2}-1.
\end{eqnarray}
The diffusion constant is reduced according to
Eq.~\eqref{eq:diff1}:
\begin{equation}
\label{eq:deff} D=D_0[1-(1+\sqrt{2}){\rm Ku}^2]=D_0(1-2.414{\rm
Ku}^2).
\end{equation}
In Fig.~\ref{fig:diff} we show the results of the numerical
simulations and compare them with our analytical result.  The
agreement with Eq.~\eqref{eq:diff1} remains very accurate up to
values of ${\rm Ku}$ around 0.25. For larger Kubo number it is
required to take into account terms of higher order in the
expansion discussed in the appendix (Eq.~\eqref{eq:app2}). As an
example, we also calculate the diffusion constant for the
correlation function $C(x,t)=\sigma^2{\rm exp}(-x^2/2\xi^2){\rm
exp}(-|t|/\tau)$:
\begin{eqnarray}
D&=&D_0(1-2.5{\rm Ku}^2),\nonumber\\
D_0&=&\sigma^2\tau.
\end{eqnarray}

\subsection{Results for the weak external force limit}
\begin{figure}[t]
\centerline{\includegraphics[width=10cm]{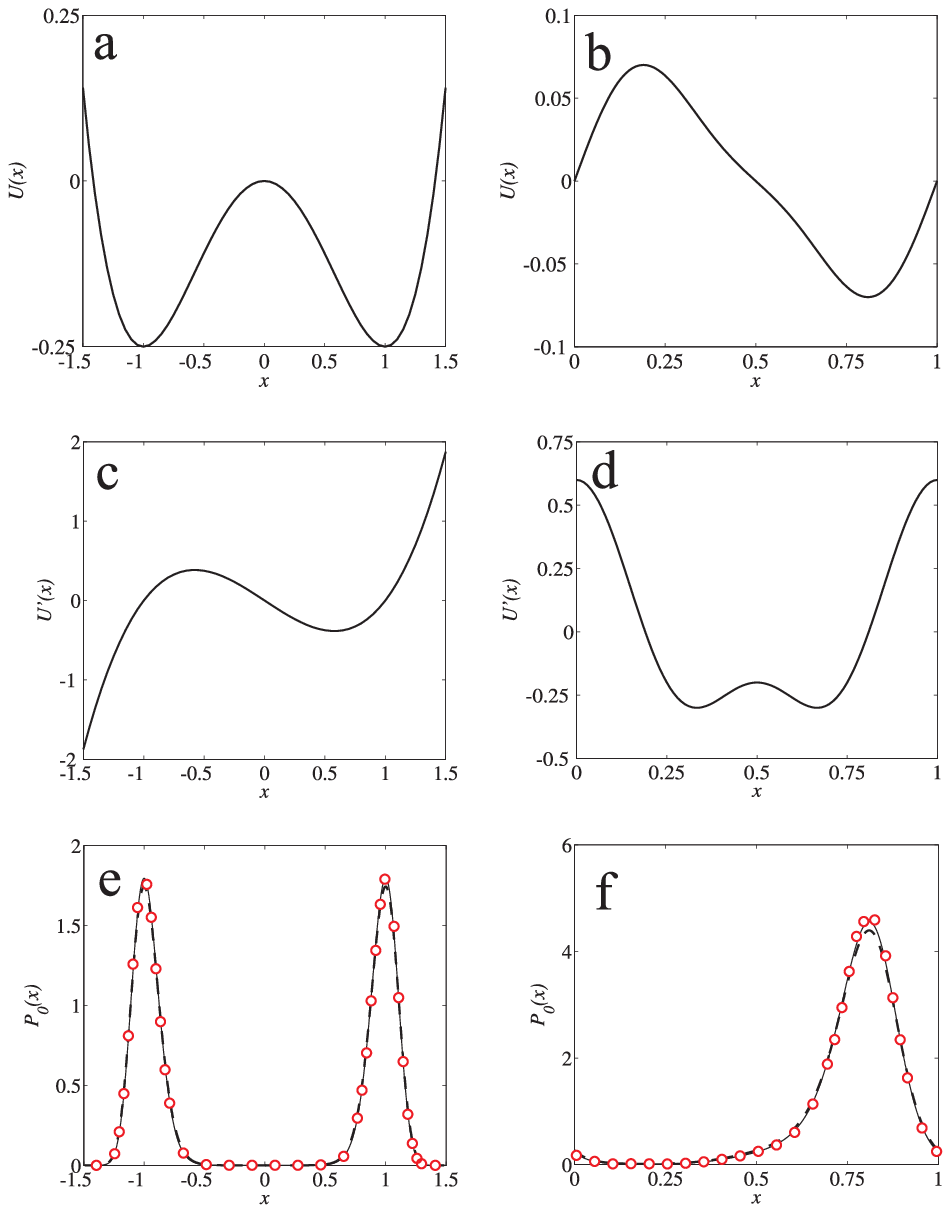}}
\caption{\label{fig:1} Probability density in the generalised
Kramers model in the limit of weak external force for small ${\rm
Ku}$. The results in panels {\bf e} and {\bf f} are for the motion
in non-periodic potential $U(x)=x^4/4-x^2/2$ ({\bf a}) and
periodic potential $U(x)=(V_0 L/2 \pi)[\sin (2\pi x/L)+ k\sin
(4\pi x/L)]$ ({\bf b}), respectively. The corresponding external
force $U'(x)$ is shown in panels {\bf c} and {\bf d}. Data from
the numerical simulations (circles) are compared with
Eq.~\eqref{eq:p01} (solid lines). Corresponding PDFs for the
classical model (dashed line) are given by Eq.~\eqref{eq:p01} with
$\alpha=0$. PDFs from the numerical simulations and theoretical
curves are normalised. Parameters of the random force are
$\sigma=1.0$, $\tau=0.02$, $\xi=0.1$, $\eta=1.0$. For the periodic
case we set $V_0=0.4$, $L=1$, and $k=0.25$.}
\end{figure}
\begin{figure}[t]
\centerline{\includegraphics[width=10cm]{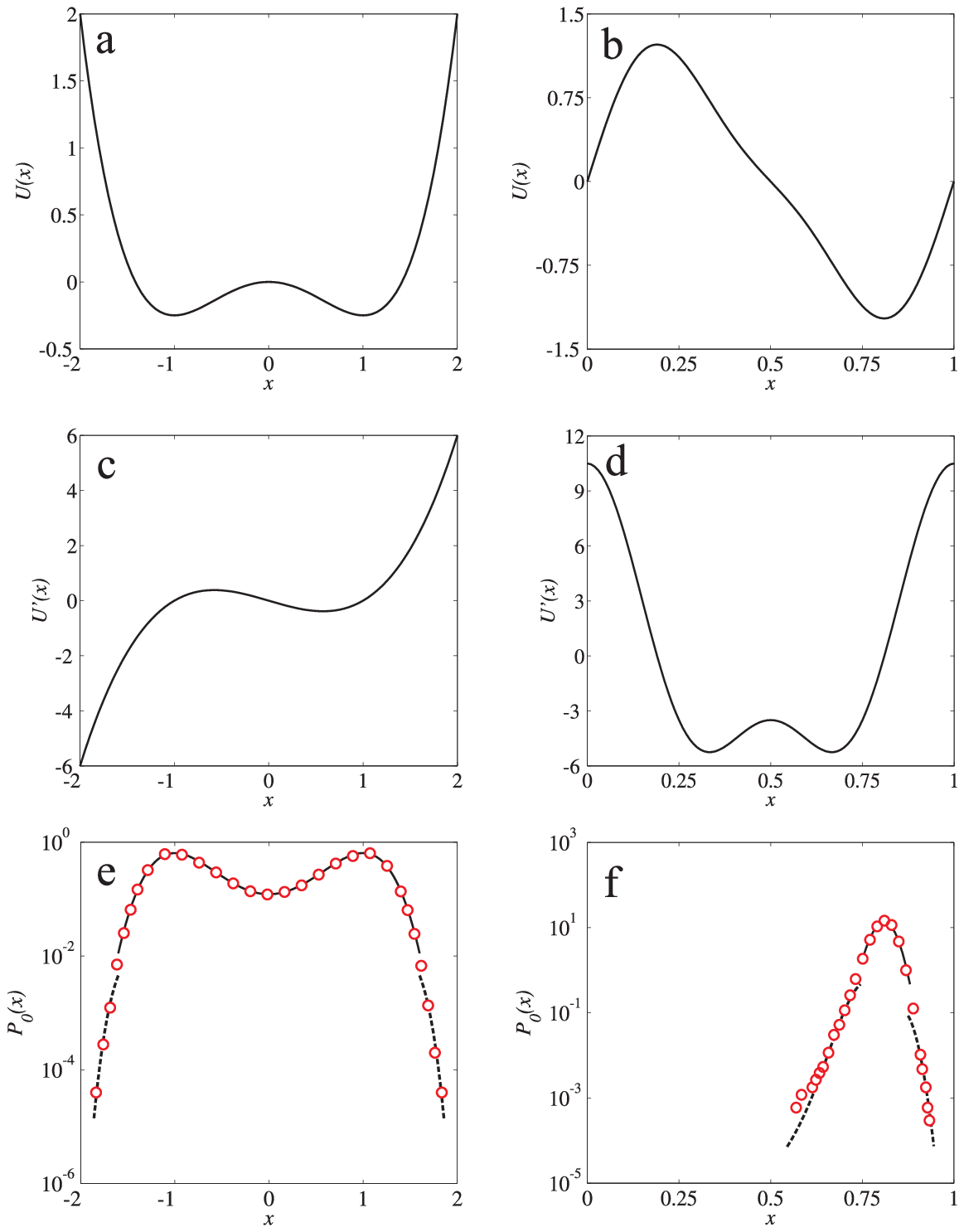}}
\caption{\label{fig:2} Probability density in the generalised
Kramers model in the limit of strong external force. The
organisation of the panels is the same as in Fig.~\ref{fig:1}.
Around the stagnation points of the potential the data from the
simulations are fitted with the Maxwellian distribution. At the
regions of large $U'(x)$ the data are compared with
Eq.~\eqref{eq:p03} and Eq.~\eqref{eq:p02} for the non-periodic and
asymmetric periodic potentials, respectively. PDFs from the
numerical simulations are normalised. Parameters of the random
force are $\sigma=2.0$, $\tau=0.05$, $\xi=0.1$, $\eta=1.0$. For
the periodic case we set $V_0=7.0$, $L=1$, and $k=0.25$.}
\end{figure}
We continue by illustrating the results in the presence of the
potential in the limit of small ${\rm Ku}$. The numerical results
and their comparison with the theory in this case are presented in
Fig.~\ref{fig:1}. For the double-well potential the particles are
concentrated around two minima of the potential with a spread
which is smaller compared to the classical Kramers model, where
the correlation length is infinite. When the Kubo number is small,
the particles almost always stays in the region of small $U'(x)$
for the double-well potential, so that the PDF is Maxwellian and
accurately given by Eq.~\eqref{eq:p01}. For the periodic potential
the value of the external force is bounded by the value of $V_0$,
which is kept sufficiently small. Although the difference between
the results for the classical Kramers model and the generalised
one is quite marginal for small ${\rm Ku}$, the tendency of the
effective increase of the potential is evident.

\subsection{Results for the strong external force limit}
Finally, we comment on the results for the strong external force
limit, summarised in Fig.~\ref{fig:2}. We remark that the
probability for the particle to propagate to the regions where
$U'(x)$ is large is typically very small. Thus, we expect our
theoretical result to give a good agreement with the tails of the
PDF far from the stagnation points of the potential. This
agreement is best seen on the logarithmic scale, as shown in
Figs.~\ref{fig:2}e and \ref{fig:2}f. Around the stagnation points
of the potential we approximate the PDF by the Maxwellian
distribution ${\rm exp}[-cU(x)]$ and choose $c$ to give the best
agreement with the data from the simulations.

The case of an asymmetric periodic potential is particularly
interesting, since it exhibits a non-zero probability flux. As we
have already discussed in section V, if the potential is
asymmetric, the particles are expected to favour a slope with a
smaller inclination to escape the minimum. The direction of the
transport in the generalised model can be deduced from
Eq.~\eqref{eq:ix}. We note that the sign of $I(L)$ is determined
by the sign of the steepest of two slopes of the potential. In
Fig.~\ref{fig:2}, it is the slope to the right of the minimum that
corresponds to $U'(x)>0$ and $U(L)>0$. From Eq.~\eqref{eq:j} we
obtain $J_0<0$, implying that it is easier for particles to escape
from the minimum using a left slope, as expected.

\begin{acknowledgments}
The author thanks Michael Wilkinson for fruitful discussions and
Markus B\"{u}ttiker for drawing attention to refs. \cite{But1987}
and \cite{Kam88}. The financial support from The Open University
is gratefully acknowledged.
\end{acknowledgments}

%\section{References}

\appendix

\section{Diffusion constant in the limit of small Kubo number}
In the appendix we discuss the problem of free diffusion in the
limit of small Kubo number. We consider a particle at the position
$x(t)$ moving with the random velocity $u(x,t)$, so that the
equation of motion reads
\begin{equation}
\dot{x}=u(x,t).
\end{equation}
The random velocity $u(x,t)$ is a stationary and translationally
invariant Gaussian random process with zero mean and correlation
function $\langle u(0,0)u(x,t)\rangle = C(x,t)$. We denote the
typical magnitude of the velocity by $u_0$ and the correlation
time and correlation length by $\tau$ and $\xi$, respectively. The
correlation function is assumed to be smooth and differentiable
and decays rapidly for $|x|>\xi$ and $|t|>\tau$. The Kubo number
measures a typical distance travelled by the particle in one
correlation time relative to the correlation length:
\begin{equation}
{\rm Ku}=\frac{u_0 \tau}{\xi}.
\end{equation}
Assuming that $x(0)=0$ the displacement after time $t$ is
\begin{equation}
x(t)=\int _0^t dt_1 \ u(x(t_1),t_1).
\end{equation}
In the limit of small ${\rm Ku}$ the spatial dependence of the
random velocity is weak, so that we can expand the trajectory in
the series as follows:
\begin{equation}
\label{eq:app2} x(t)\approx \int _0^t dt_1 \ u(0,t_1)+\int _0^t
dt_1 \ \frac{\partial u(0,t_1)}{\partial x}x(t_1) + \frac{1}{2}
\int _0^{t} dt_1 \ \frac{\partial ^2 u(0,t_1)}{\partial x^2}
x^2(t_1).
\end{equation}
This expansion includes all the terms which will yield only two-
and four-point correlation functions. After squaring and averaging
Eq.~\eqref{eq:app2} we obtain
\begin{eqnarray}
\langle x^2(t)\rangle &\approx &\int _0^t dt_1 \int _0^t dt_2 \
\langle u(0,t_1)u(0,t_2)\rangle +\int _0^t dt_1 \int _0^t dt_2 \
\left\langle \frac{\partial u(0,t_1)}{\partial x}\frac{\partial
u(0,t_2)}{\partial x} x(t_1)x(t_2)\right \rangle \nonumber \\
&+& 2\int _0^t dt_1 \int _0^t dt_2 \ \left\langle
u(0,t_1)\frac{\partial u(0,t_2)}{\partial x} x(t_2)\right
\rangle\nonumber\\&+&\int _0^t dt_1 \int _0^t dt_2  \ \left\langle
u(0,t_1)\frac{\partial^2 u(0,t_2)}{\partial x^2} x^2(t_2)\right
\rangle.
\end{eqnarray}
Expanding $x(t_1)$ and $x(t_2)$ further we obtain
\begin{eqnarray}
\label{eq:app3} \langle x^2(t)\rangle &\approx &\int _0^t dt_1
\int _0^t dt_2 \ \langle u(0,t_1)u(0,t_2)\rangle
\nonumber\\&+&\int _0^t dt_1 \int _0^t dt_2 \int _0^{t_1} dt' \int
_0^{t_2} dt'' \ \left\langle \frac{\partial u(0,t_1)}{\partial
x}\frac{\partial
u(0,t_2)}{\partial x} u(0,t')u(0,t'')\right \rangle \nonumber \\
&+& 2\int _0^t dt_1 \int _0^t dt_2 \int _0^{t_2} dt' \int _0^{t'}
dt'' \ \left\langle u(0,t_1)\frac{\partial u(0,t_2)}{\partial x}
\frac{\partial u(0,t')}{\partial x}u(0,t'')\right
\rangle\nonumber\\
&+&\int _0^t dt_1 \int _0^t dt_2 \int _0^{t_2} dt' \int _0^{t_2}
dt'' \ \left\langle u(0,t_1)\frac{\partial^2 u(0,t_2)}{\partial
x^2} u(0,t')u(0,t'')\right \rangle.
\end{eqnarray}
Here, the first term corresponds to the motion of the particle in
the absence of spatial dependence of the velocity. Assuming that
$t\gg \tau$ we obtain
\begin{equation}
\int _0^t dt_1 \int _0^t dt_2 \ \langle
u(0,t_1)u(0,t_2)\rangle\approx t\int _{-\infty}^{\infty} dt \
C(0,t)=2D_0t,
\end{equation}
where
\begin{equation}
D_0=\frac{1}{2} \int _{-\infty}^{\infty} dt \ C(0,t) \sim
u_0^2\tau.
\end{equation}
For the rest of the terms in Eq.~\eqref{eq:app3} we use the
following statistical properties of a Gaussian noise:
\begin{eqnarray}
\langle u(0,t_1)u(0,t_2)u(0,t_3)u(0,t_4)\rangle
&=&C(0,t_1-t_2)C(0,t_3-t_4)\nonumber \\&+&C(0,t_1-t_3)C(0,t_2-t_4)\nonumber \\&+&C(0,t_1-t_4)C(0,t_2-t_3),\nonumber\\
\left\langle \frac{\partial u(0,t_1)}{\partial x} \frac{\partial
u(0,t_2)}{\partial x} \right \rangle &=& -\frac{\partial^2
C(0,t_1-t_2)}{\partial x^2},\nonumber\\
\left\langle \frac{\partial^2 u(0,t_1)}{\partial x^2} u(0,t_2)
\right \rangle &=& \frac{\partial^2 C(0,t_1-t_2)}{\partial x^2},\nonumber\\
\left\langle \frac{\partial u(0,t_1)}{\partial x} u(0,t_2) \right
\rangle &=& 0.
\end{eqnarray}
In the discussion below we shall drop the spatial argument of the
correlation function implying that $C(t)\equiv C(0,t)$. We have
\begin{eqnarray}
\label{eq:app5} &&\langle x^2(t)\rangle \approx 2D_0t-\int _0^t
dt_1 \int _0^t dt_2 \int _0^{t_1} dt' \int _0^{t_2} dt'' \
\frac{\partial^2
C(t_1-t_2)}{\partial x^2}C(t'-t'')\nonumber \\
&-& 2\int _0^t dt_1 \int _0^t dt_2 \int _0^{t_2} dt' \int _0^{t'}
dt'' \ \frac{\partial^2 C(t_2-t')}{\partial x^2} C(t_1-t'')\nonumber\\
&+&\int _0^t dt_1 \int _0^t dt_2 \int _0^{t_2} dt' \int _0^{t_2}
dt'' \nonumber\\
&\times &\left[\frac{\partial^2 C(t_1-t_2)}{\partial x^2}
C(t'-t'')+\frac{\partial^2 C(t_2-t'')}{\partial x^2}
C(t_1-t')+\frac{\partial^2 C(t_2-t')}{\partial x^2}
C(t_1-t'')\right].
\end{eqnarray}
We note that the following relation holds for any $a$ and the
correlation function which can be written in the form
$C(x,t)=C_x(x)C_t(t)$:
\begin{equation}
\int _0^{a} dt' \int _0^{a} dt'' \ \frac{\partial^2
C(t_2-t'')}{\partial x^2} C(t_1-t')=\int _0^{a} dt' \int _0^{a}
dt'' \ \frac{\partial^2 C(t_2-t')}{\partial x^2} C(t_1-t'').
\end{equation}
Using this we can combine the terms in Eq.~\eqref{eq:app5}
together and obtain
\begin{eqnarray}
\label{eq:app1} \langle x^2(t)\rangle &\approx & 2D_0t+\int _0^t
dt_1 \int _0^t dt_2  \ \frac{\partial^2 C(t_1-t_2)}{\partial x^2}
\int _{t_1}^{t_2} dt'
\int _{0}^{t_2} dt'' \ C(t'-t'')\nonumber \\
&+& 2\int _0^t dt_1 \int _0^t dt_2 \int _0^{t_2} dt' \int
_{t'}^{t_2} dt'' \ \frac{\partial^2 C(t_2-t')}{\partial x^2}
C(t_1-t'').
\end{eqnarray}
We now calculate the remaining two terms. We first consider
\begin{equation}
Q_1=\int _0^t dt_1 \int _0^t dt_2 \ \frac{\partial^2
C(t_1-t_2)}{\partial x^2} \int _{t_1}^{t_2} dt' \int _{0}^{t_2}
dt'' \ C(t'-t'').
\end{equation}
%We surmise that this term is linear in $t$ and therefore the
%integral in $t'$ and $t''$ depends only on $t_2-t_1$.
We put $T_1=t'-t_2$ and $T_2=t'-t''$ to obtain
\begin{equation}
Q_1=\int _0^t dt_1 \int _0^t dt_2 \ \frac{\partial^2
C(t_1-t_2)}{\partial x^2} \int _{t_1-t_2}^{0} dT_1 \int
_{T_1}^{T_1+t_2} dT_2 \ C(T_2).
\end{equation}
We note that $C(T_2)$ is significant around $T_2=0$ for $t\gg
\tau$. We have
\begin{equation}
Q_1\approx \int _0^t dt_1 \int _0^t dt_2 \ \frac{\partial^2
C(t_1-t_2)}{\partial x^2} \int _{t_1-t_2}^{0} dT_1 \int
_{T_1}^{\infty} dT_2 \ C(T_2).
\end{equation}
Now, we denote $T_3=t_1-t_2$ and obtain
\begin{equation}
Q_1\approx t \int _{-\infty}^{\infty} dT_3 \ \frac{\partial^2
C(T_3)}{\partial x^2} \int _{T_3}^{0} dT_1 \ \int _{T_1}^{\infty}
dT_2 \ C(T_2)
\end{equation}
or
\begin{equation}
Q_1\approx \beta_1 t,
\end{equation}
where
\begin{equation}
\label{eq:beta1} \beta_1=\int _{-\infty}^{\infty} dt_1 \
\frac{\partial^2 C(t_1)}{\partial x^2} \int _{t_1}^{0} dt_2 \int
_{t_2}^{\infty} dt_3 \ C(t_3).
\end{equation}
We remark that $\beta_1\sim u_0^4\tau^3/\xi^2$ or, equivalently,
$\beta_1\sim D_0{\rm Ku}^2$.

We now consider the third term in Eq.~\eqref{eq:app1}:
\begin{equation}
Q_2=\int _0^t dt_1 \int _0^t dt_2 \int _0^{t_2} dt' \int
_{t'}^{t_2} dt'' \  \frac{\partial^2 C(t_2-t')}{\partial x^2}
C(t_1-t'').
\end{equation}
We introduce new variables $T_1=t_2-t'$ and $T_2=t_1-t''$ and
obtain
\begin{equation}
Q_2=\int _0^t dt_1 \int _0^t dt_2 \int _0^{t_2} dT_1 \int
_{t_1-t_2}^{t_1-t_2+T_1} dT_2 \ \frac{\partial^2 C(T_1)}{\partial
x^2} C(T_2).
\end{equation}
The integrand here is significant around $T_1=0$, so that we can
write
\begin{equation}
Q_2\approx \int _0^t dt_1 \int _0^t dt_2 \int _0^{\infty} dT_1
\int _{t_1-t_2}^{t_1-t_2+T_1} dT_2 \ \frac{\partial^2
C(T_1)}{\partial x^2} C(T_2).
\end{equation}
We denote $T_3=t_1-t_2$ and obtain
\begin{equation}
Q_2\approx \int _0^t dt_1 \int _{t_1-t}^{t_1} dT_3 \int
_0^{\infty} dT_1 \int _{T_3}^{T_3+T_1} dT_2 \ \frac{\partial^2
C(T_1)}{\partial x^2} C(T_2).
\end{equation}
For $t\gg \tau$ we have
\begin{equation}
Q_2\approx t \int _{-\infty}^{\infty} dT_3 \int _0^{\infty} dT_1
\int _{T_3}^{T_3+T_1} dT_2 \ \frac{\partial^2 C(T_1)}{\partial
x^2} C(T_2)
\end{equation}
or
\begin{equation}
Q_2\approx \beta_2 t,
\end{equation}
where
\begin{equation}
\beta_2=\int _{-\infty}^{\infty} dt_1 \int _0^{\infty} dt_2 \int
_{t_1}^{t_1+t_2} dt_3 \ \frac{\partial^2 C(t_2)}{\partial x^2}
C(t_3).
\end{equation}
We note that the integrand is significant around $t_2=0$ so that
we can write
\begin{equation}
\int _{t_1}^{t_1+t_2} dt_3 \  C(t_3)\approx C(t_1)t_2
\end{equation}
and therefore
\begin{equation}
\beta_2=\int _{-\infty}^{\infty} dt_1 \ C(t_1) \int _0^{\infty}
dt_2 \ t_2 \ \frac{\partial^2 C(t_2)}{\partial x^2} =-2D_0\alpha,
\end{equation}
where $\alpha>0$ is defined similarly to Eq.~\eqref{eq:alpha}:
\begin{equation}
\alpha=-\int _0^{\infty} dt \ t \ \frac{\partial^2 C(t)}{\partial
x^2}.
\end{equation}
We remark that $\alpha\sim {\rm Ku}^2$. Going back to
Eq.~\eqref{eq:app1} we obtain
\begin{equation}
\langle x^2(t)\rangle \approx 2D_0t+\beta_1 t + 2\beta_2 t.
\end{equation}
We may define the effective diffusion constant
$D_{eff}=D_0+\beta_1/2+\beta_2$, so that the mean square
displacement can be written as
\begin{equation}
\langle x^2(t)\rangle \approx 2D_{eff}t.
\end{equation}
In the view that $\beta_1\sim D_0\alpha$, we may write
$\beta_1=-2\gamma\alpha D_0$ for some factor $\gamma>0$ which can
be expressed from Eq.~\eqref{eq:beta1}. The effective diffusion
constant is then given by
\begin{equation}
D_{eff}=D_0[1-\alpha(2+\gamma)].
\end{equation}
We remark that both $\beta_1$ and $\beta_2$ are negative for
$C(x,t)=C_x(x)C_t(t)$ and $C_t(t)>0$. Thus, the diffusion constant
is reduced and the reduction is proportional to ${\rm Ku}^2$ when
${\rm Ku}$ is small.

%As expected, the diffusion constant is reduced, because in the
%presence of spatial dependence the force experienced by the moving
%particle de-correlates more rapidly than for a stationary
\end{document}